\documentclass[12pt]{article}
\usepackage[dvips]{graphicx}

\begin{document}

\title{Optical theorem and unitarity}
\author{V.I.Nazaruk\\
Institute for Nuclear Research of RAS, 60th October\\
Anniversary Prospect 7a, 117312 Moscow, Russia.*}

\date{}
\maketitle
\bigskip

\begin{abstract}
It is shown that an application of optical theorem for the non-unitary $S$-matrix can lead to the qualitative error in the result.
\end{abstract}

\vspace{5mm}

\vspace{5mm}

\vspace{1cm}

*E-mail: nazaruk@inr.ru

\newpage
\setcounter{equation}{0}
\section{Introduction}
The importance of the unitarity condition is well known [1,2]. Optical theorem should be applied for unitary $S$-matrix. It is frequently used for the non-Hermitian Hamiltonians as well. In this case the $S$-matrix should be unitarized. However, this requirement breaks down for a number of well-known models. The possible consequences are demonstrated in this paper.

We recall that unitarity condition
\begin{equation}
(SS^+)_{fi}=\delta _{fi},
\end{equation}
$S=1+iT$, gives
\begin{equation}
2ImT_{ii}=\sum_{f}\mid T_{fi}\mid ^2.
\end{equation}
>From this equation the optical theorem and expression for the decay width 
\begin{equation}
\Gamma _{opt}=\frac{1}{T_0}(1-\mid S_{ii}\mid ^2)\approx \frac{1}{T_0}2ImT_{ii}
\end{equation}
are obtained. Here $T_0$ is the normalization time, $T_0\rightarrow \infty $.
The non-unitarity of $S$-matrix implies that $(SS^+)_{fi}\neq \delta _{fi}$ or, what is the same
\begin{equation}
(SS^+)_{fi}=\delta _{fi}+\alpha _{fi},
\end{equation}
$\alpha _{fi}\neq 0$, resulting in 
\begin{equation}
2ImT_{ii}=\sum_{f}\mid T_{fi}\mid ^2-\alpha _{fi}\neq \sum_{f}\mid T_{fi}\mid ^2
\end{equation}		
since the value $\sum_{f}\mid T_{fi}\mid ^2$ can be very small. Instead of (2) we have (5) and eq. (3) and optical theorem are inapplicable. Also eq. (4) means the probability non-conservation: $\sum_{f}\mid S_{fi}\mid ^2\neq 1$. 

\section{Calculation}
We demonstrate the consequences of the incorrect application of eqs. (2) and (3) for the non-unitary $S$-matrix by the example of the $n\bar{n}$ transition in a medium followed by annihilation [3-5]:
\begin{equation}
n\rightarrow \bar{n}\rightarrow M.
\end{equation}
Here $M$ are the annihilation mesons (see Fig.1). This is a simplest process which allows a result in the analytical form for unitary and non-unitary models.

The background potential of neutron-medium interaction $U_n$ is included in the neutron wave function: $n(x)=\Omega ^{-1/2}\exp (-i\epsilon _nt+i{\bf p}_n{\bf x})$, $\epsilon _n={\bf p}_n^2/2m+U_n$. The interaction Hamiltonian is
\begin{equation}
{\cal H}_I={\cal H}_{n\bar{n}}+{\cal H},
\end{equation}
\begin{equation}
{\cal H}_{n\bar{n}}=\epsilon \bar{\Psi }_{\bar{n}}\Psi _n+H.c.,
\end{equation}
\begin{equation}
{\cal H}=V\bar{\Psi }_{\bar{n}}\Psi_{\bar{n}}+{\cal H}_a.
\end{equation}
Here ${\cal H}_{n\bar{n}}$ and ${\cal H}$ are the Hamiltonians of $n\bar{n}$ conversion [6,7] and $\bar{n}$-medium interaction, respectively; ${\cal H}_a$ and $V$ are the effective annihilation Hamiltonian and the residual scalar field, respectively; $\epsilon $ is a small parameter [6].

\subsection{Models with the Hermitian Hamiltonian}
Let us the amplitude of antineutron annihilation in the medium $M_a$ is totally dependent on a Hamiltonian ${\cal H}_a$:
\begin{equation}
<\!f0\!\mid T\exp (-i\int dx{\cal H}_a(x))-1\mid\!0\bar{n}_{p}\!>=
N(2\pi )^4\delta ^4(p_f-p_i)M_a.
\end{equation}
Here $\mid\!0\bar{n}_{p}\!>$ is the state of the medium containing the $\bar{n}$ with the 4-momentum $p=(\epsilon ,{\bf p})$; $<\!f\!\mid $ denotes the annihilation products, $N$ includes the normalization factors of the wave functions. The antineutron annihilation width $\Gamma $ is expressed 
through $M_a$: $\Gamma \sim \int d\Phi \mid\!M_a\!\mid ^2$.		

\begin{figure}[h1]
%  \reflectbox{\includegraphics[height=.3\textheight]{golfer}}
  {\includegraphics[height=.25\textheight]{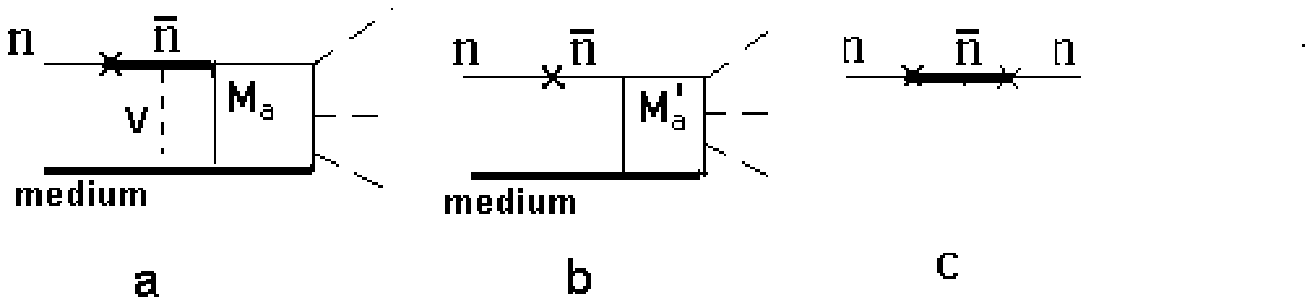}}
  \caption{{\bf a} $n\bar{n}$ transition in the medium followed by annihilation.  {\bf b} Same as {\bf a} but the annihilation amplitude is given by (16). {\bf c}  The on-diagonal matrix element $T_{ii}$ (see text)}
\end{figure}

In the lowest order in ${\cal H}_{n\bar{n}}$ the amplitude of process (6) is {\em uniquely} determined by the Hamiltonians (7)-(9) (see Fig. 1a):
\begin{equation}
M=\epsilon G_VM_a,
\end{equation}
where the antineutron Green function $G_V$ is
\begin{equation}
G_V=G+GVG+...=\frac{1}{(1/G)-V}=-\frac{1}{V},
\end{equation}
\begin{equation}
G=\frac{1}{\epsilon _{\bar{n}} -{\bf p}_{\bar{n}}^2/2m-U_n+i0} \sim \frac{1}{0},
\end{equation}
since ${\bf p}_{\bar{n}}={\bf p}$, $\epsilon _{\bar{n}}=\epsilon $. $G_V$ depends on $V$ only because the Hamiltonian ${\cal H}_a$ involved in $M_a$.

For the total process width $\Gamma _t$ one obtains
\begin{equation}
\Gamma _t=N_1\int d\Phi \mid\!M\!\mid ^2=\frac{\epsilon ^2}{V^2}N_1\int d\Phi \mid\!M_a\!\mid ^2=\frac{\epsilon ^2}{V^2}\Gamma ,
\end{equation}
\begin{equation}
\Gamma =N_1\int d\Phi \mid\!M_a\!\mid ^2,
\end{equation}
where  $\Gamma $ is the annihilation width of $\bar{n}$. The normalization multiplier $N_1$ is the same for $\Gamma _t$ and $\Gamma $.

If $M_a$ is determined by (10), the process amplitude (11) follows uniquely from (7)-(9). On the other hand, for the one-step process of the antineutron annihilation in the medium $(\bar{n}-\mbox{medium})\rightarrow (\mbox{annihilation products}-\mbox{medium})$, the annihilation amplitude $M'_a$ can be defined through the Hamiltonian ${\cal H}$ and not ${\cal H}_a$:
\begin{equation}
<\!f0\!\mid T\exp (-i\int dx{\cal H}(x))-1\mid\!0\bar{n}_{p}\!>=
N(2\pi )^4\delta ^4(p_f-p_i) M'_a.
\end{equation}
$M'_a$ contains all $\bar{n}$-medium interactions including antineutron rescattering in the initial state. In this case the amplitude of process (6) is
(see Fig. 1b)
\begin{equation}
M'=\epsilon GM'_a.
\end{equation}
The definition of annihilation amplitude through eq. (16) is natural since it corresponds to the observable values. There are many physical arguments in support of the model (17). However, this model contains infrared singularity
$M'\sim 1/0$ since $G\sim 1/0$. This problem has been considered in [9,10]. For the purposes of this paper it is essential that model (17) gives linier $\Gamma $-dependence $\Gamma '_t\sim \int d\Phi \mid\!M'\!\mid ^2\sim \Gamma $, as well as model (11). Consequently, the model with Hermitian Hamiltonian gives linear $\Gamma $-dependence at all definition of annihilation amplitude. 

\subsection{Model with the non-Hermitian Hamiltonian}
On the other hand, we consider the standard model of the process (6) [6-10]. In this model the $\bar{n}$-medium interaction is described by optical potential. The interaction Hamiltonian is given by (7), where
\begin{equation}
{\cal H}\rightarrow  {\cal H}_{opt}=(U_{\bar{n}}-U_n)\bar{\Psi }_{\bar{n}}\Psi_
{\bar{n}}=(V-i\Gamma /2)\bar{\Psi }_{\bar{n}}\Psi_{\bar{n}},
\end{equation}
where $U_{\bar{n}}$ is the antineutron optical potential. In eq. (18) we have put ${\rm Re} U_{\bar{n}}-U_n=V$, ${\rm Im} U_{\bar{n}}=-\Gamma /2$.

The full in-medium antineutron propagator $G_m$ is
\begin{equation}
G_m=\frac{1}{\epsilon _{\bar{n}} -{\bf p}_{\bar{n}}^2/2m-U_{\bar{n}}+i0}.
\end{equation}

The on-diagonal matrix element $T_{ii}$ is shown in Fig. 1c. For the total decay width $\Gamma _{opt}$ eq. (3) gives the well-known result [6-8]:
\begin{equation}
\Gamma _{opt}=-2{\rm Im}\epsilon G_m\epsilon =2\epsilon ^2\frac{\Gamma /2}{V^2+(\Gamma /2)^2}\approx 4\epsilon ^2/\Gamma .
\end{equation} 

\section{Summary and conclusion}
The $\Gamma $-dependence of the results differs fundamentally: $\Gamma _{opt}\sim 1/\Gamma $, whereas $\Gamma _t\sim \Gamma $. At the same time the annihilation is the main effect which defines the process speed. One of two models is wrong. 

We assert that model with non-Hermitian Hamiltonian ${\cal H}_{opt}$ is wrong since (20) follows from (3) which is inapplicable for non-unitary $S$-matrix (see (5)). We believe that the Hamiltonian (18) describes correctly the $n\bar{n}$ transition with $\bar{n}$ in the final state $(n-\mbox{medium})\rightarrow (\bar{n}-\mbox{medium})$ since eq. (3) is not used in this case.

Notice that the result (20) takes place in the all standard calculations (see, 
for example, refs. [6-8]) because they are based on the optical potential. Finally, we compare (14) and (20):
\begin{equation}
r=\frac{\Gamma _t}{\Gamma _{opt}}=\frac{\Gamma ^2}{4V^2}.
\end{equation}
For the $n\bar{n}$ transition in the nuclear matter the realistic set of parameters is $\Gamma =100$ MeV, and $V=10$ MeV. Then $r=25$. For the model (17), or else $V=0$, eq. (14) is invalid. However, in this case $r\gg 1$ as well [10].

Optical theorem can be applied only if $S$-matrix is unitary or unitarized. The same is true for the condition of probability conservation. If the optical potential is used for the problem described by Schrodinger-type equations (optical model), the unitarization takes place: optical potential is fitted to $\bar{p}$-atom ($\pi ^-$-atom, $K^-$-atom) and low energy scattering data. The optical potential is the effective one. The above-considered problem is described by the system of coupled equations. There is no experimental data and unitarization in this case. The consequences are illustrated by eqs. (14), (20) and (21).

\end{document}